\documentclass[12pt]{article}
\usepackage{epsfig}
\usepackage{amssymb}

\begin{document}

\newcommand{\beq}{\begin{equation}}
\newcommand{\eeq}{\end{equation}}
\newcommand{\barr}{\begin{eqnarray}}
\newcommand{\earr}{\end{eqnarray}}

\newcommand{\andy}[1]{ }

%%%%%%%%%%%%%%%%%%%%%%%%new def%%%%%%%%%%%%%%%%%%%%%%%%%%%
\newcommand{\ket}[1]{| #1 \rangle}
\newcommand{\bra}[1]{\langle #1 |}
%%%%%%%%%%%%%%%%%%%%%%%%%%%%%%%%%%%%%%%%%%%%%%%%%%%%%%%%%%%

\begin{titlepage}

\vspace{.5cm}
\begin{center}
{\LARGE Decoherence and fluctuations in quantum interference experiments

}

\quad

{\large A. Mariano, P. Facchi and S. Pascazio
\\
           \quad \\
Dipartimento di Fisica, Universit\`a di Bari \\
and Istituto Nazionale di Fisica Nucleare, Sezione di Bari \\
 I-70126 Bari, Italy \\
}

\vspace*{.5cm}

{\small\bf Abstract}\\ \end{center}

{\small We analyze the notion of quantum coherence in an
interference experiment. We let the phase shifts fluctuate
according to a given statistical distribution and introduce a
decoherence parameter, defined in terms of  a generalized
visibility of the interference pattern. One might naively expect
that a particle ensemble suffers a greater loss of quantum
coherence by interacting with an increasingly randomized
distribution of shifts. As we shall see, this is not always true.
}

\vspace*{.5cm} PACS: 03.75.Dg, 05.40.-a, 42.50.-p

\end{titlepage}

\newpage

\section{Introduction}
\label{sec-int}\andy{sec-int}

Decoherence is an interesting phenomenon and a topic that attracts
widespread attention \cite{Dec}. However, it is not easy to give a
quantitative definition of decoherence \cite{fibonacci}. All
attempts at defining it always depend on the experimental
configuration and on the authors' taste. An interesting related
quantity is the square of the density matrix \cite{Watanabe}. This
quantity enjoys interesting features
\cite{Manfredi}, but also yields results which are at variance
with naive expectations based on entropy \cite{fibonacci}. We
consider here an alternative, operational definition of
decoherence, based on a fluctuation approach \cite{sim}, and
discuss its physical meaning by considering some examples.

\section{Fluctuating phase shifter}
\label{sec-asm}\andy{sec-asm}
Consider a Mach-Zender interferometer (MZI), with a phase shifter
$\Delta$ in one of its two arms. If $\ket{\psi_{\rm in}}$ is the
incoming state, the output state in the ordinary channel is
\andy{uscint}
\beq
\label{eq:uscint}
\ket{\psi_{O}}=\frac{1}{2}\left[1+
e^{\frac{i}{\hbar}\hat p\Delta}\right]\ket{\psi_{\rm
in}}\equiv\hat O(\Delta)\ket{\psi_{\rm in}}
\eeq
or, in terms of the density matrix,
\andy{matde}
\beq
\label{eq:matde}
\hat\rho_{O}\equiv\ket{\psi_{
O }}\bra{\psi_{O }}=\hat O(\Delta)\ket{\psi_{\rm
in}}\bra{\psi_{\rm in}}
\hat
O(\Delta)^\dagger=\hat O(\Delta)\hat{\rho_{\rm in}}\hat O(\Delta)^\dagger,
\eeq
where $\hat\rho_{\rm in}$ is the density matrix of the incoming
state. Suppose now that the shift $\Delta$ fluctuates according to
a probability function $w(\Delta-\Delta_0)$. The trace of the
average density matrix is
\andy{tmadem}
\beq\label{eq:tmadem}
{\rm Tr }\;\overline{\hat\rho_{O}}={\rm Tr
}\;\int\;d\Delta\;w(\Delta-\Delta_0) \hat O(\Delta)\hat \rho_{\rm in}\hat
O(\Delta)^\dagger={\rm Tr }\;\left(\hat \rho_{\rm in}\overline{\hat
O(\Delta)^\dagger\hat O(\Delta)}\right)
\eeq
and one obtains, after some algebra,
\andy{oocro}
\beq\label{eq:oocro}
\overline{\hat O(\Delta)^\dagger\hat
O(\Delta)}=\frac{1}{2}\left(1+\overline{\cos\frac{\hat
p\Delta}{\hbar}}\right).
\eeq
Consider now the Fourier transform of the probability density of
the fluctuations
\andy{wtilde}
\barr\label{eq:wtilde}
\Omega(\hat
p)\equiv\int\;d\Delta\; w(\Delta) e^{\frac{i}{\hbar}\hat p\Delta}
=\int\;d\Delta\;w(\Delta)\cos\frac{\hat
p\Delta}{\hbar}+i\int\;d\Delta\; w(\Delta)\sin\frac{\hat
p\Delta}{\hbar}.
\earr
If we assume that the distribution of fluctuations is symmetric,
$w(\Delta)=w(-\Delta)$, we get
\andy{copdel}
\barr\label{eq:copdel}
\overline{\cos\frac{\hat p \Delta}{\hbar}
}=\int\;d\Delta\;w(\Delta-\Delta_0)\cos\frac{\hat p \Delta }{\hbar
} =\cos\frac{\hat p\Delta_0 }{\hbar }\;\Omega(\hat p)
\earr
and (\ref{eq:oocro}) becomes
\andy{omoco}
\beq\label{eq:omoco}
\overline{\hat O(\Delta)^\dagger\hat
O(\Delta)}=\frac{1}{2}\left[1+\Omega(\hat p)\cos\frac{\hat
p\Delta_0 }{\hbar }\right].
\eeq
We notice that the same results are obtained with a different
setup: consider a polarized neutron that interacts with a magnetic
field perpendicular to its spin. Due to the longitudinal
Stern-Gerlach effect, its wave packet is split into two components
that travel with different speed and are therefore separated in
space
\cite{BRSW}. After a projection onto the initial spin state, the
final state reads
\andy{finmag}
\beq
\label{eq:finmag}
\ket{\psi_{\|}}=\frac{1}{2}\left[
e^{-\frac{i}{2\hbar}\hat p\Delta}+e^{\frac{i}{2\hbar}\hat
p\Delta}\right]\ket{\psi_{\rm in}}\equiv\hat
O'(\Delta)\ket{\psi_{\rm in}},
\eeq
where $\Delta$ is in this case the spatial separation between the
two wave packets corresponding to the two spin components. By
averaging over $\Delta$ it is easy to show that one obtains again
(\ref{eq:omoco}).

By plugging the average operator (\ref{eq:omoco}) into
(\ref{eq:tmadem}) one finally gets
\andy{fintr}
\beq\label{eq:fintr}
{\rm Tr }\;\overline{\hat\rho_O
}=\frac{1}{2}\left[1+\left\langle\Omega(\hat p)\cos\frac{\hat
p\Delta_0 }{\hbar }\right\rangle\right],
\eeq
where $\langle\cdots\rangle$ denotes the expectation value over
state $\rho_{\rm in}$. On the other hand, the momentum
distribution is easily proved to read
\andy{imptr}
\beq\label{eq:imptr}
P_O(p)=\bra p\overline{\hat\rho_O }\ket p =\frac{1}{2}P_{\rm in}(p)
\left[1+\Omega(p)\cos\frac{p\Delta_0}{\hbar }\right],
\eeq
where
\andy{imptr0}
\beq\label{eq:imptr0}
P_{\rm in}(p)=\bra p\hat \rho_{\rm in}\ket p, \qquad
\Omega(p)=\bra p \Omega(\hat p)\ket p.
\eeq
We now introduce the {\em visibility} of the interference pattern
\andy{visibp}
\beq\label{eq:visibp}
{\cal V}(p)\equiv\frac{P_O(p)_{\rm MAX }-P_O(p)_{\rm min}
}{P_O(p)_{\rm MAX }+P_O(p)_{\rm min} }=|\Omega(p)|,
\eeq
where $P_O(p)_{\rm MAX }$ ($P_O(p)_{\rm min}$) is the maximum
(minimum) value assumed by $P_O(p)$ when $\Delta_0$ varies. Notice
that, according to this definition, the visibility is a function
of momentum $p$. By using (\ref{eq:wtilde}) and (\ref{eq:imptr0}),
one infers that the visibility is the modulus of the Fourier
transform of the distribution of the shifts $\Delta$ and is
therefore a quantity which is closely related to the physical
features of the phase shifter. We now turn to a definition of
decoherence.

\section{Operational definition of decoherence}
\label{sec-cont} \andy{sec-cont}
Consider the MZI introduced in the previous section. The relative
frequency of particles detected in the ordinary channel is, by
(\ref{eq:fintr}),
\andy{ordpar}
\beq\label{eq:ordpar}
{\cal N}_O(\Delta_0)={\rm Tr }\;\overline{\hat\rho_O }
=\frac{1}{2}\left[1+\left\langle\Omega(\hat p)\cos\frac{\hat
p\Delta_0}{\hbar}\right\rangle\right].
\eeq
On the other hand, in the extraordinary ($E$) channel we get
\andy{strpar}
\beq\label{eq:strpar}
{\cal N}_E(\Delta_0)={\rm Tr }\;\overline{\hat\rho_E}
=\frac{1}{2}\left[1-\left\langle\Omega(\hat p)\cos\frac{\hat
p\Delta_0}{\hbar}\right\rangle\right].
\eeq
(Note that ${\cal N}_O+{\cal N}_E=1$.) Their difference is
\andy{de0vis}
\beq\label{eq:de0vis}
{\cal N}_O(\Delta_0)-{\cal N}_E(\Delta_0)=
\left\langle\Omega(\hat p)\cos\frac{\hat
p\Delta_0}{\hbar}\right\rangle =\int\;dp\;P_{\rm
in}(p)\Omega(p)\cos\frac{p\Delta_0 }{\hbar }
\eeq
and one can define a {\em generalized visibility}
\andy{de0visb}
\beq\label{eq:de0visb}
{\cal V}=\max_{\Delta_0 }|{\cal N}_O(\Delta_0)-{\cal
N}_E(\Delta_0)| =\max_{\Delta_0 }\left|\left\langle
\Omega(\hat p)\cos\frac{\hat p\Delta_0
}{\hbar }\right\rangle\right|.
\eeq
Notice that when $P_{\rm in}(p')= \delta(p'-p)$ ({\em normalized}
monochromatic incoming state, to be improperly referred to as
plane wave of momentum $p$), the generalized visibility reduces to
the standard ``local" visibility (\ref{eq:visibp})
\andy{qwe}
\barr\label{eq:qwe}
{\cal V } = \max_{\Delta_0}
\left|\int\;dp'\;\delta(p'-p)\Omega(p')\cos\frac{p'\Delta_0}{\hbar
} \right| = \max_{\Delta_0}
\left|\Omega(p)\;\cos\frac{p\Delta_0}{\hbar}\right|={\cal V}(p) .
\earr
In general one gets
\andy{rty}
\beq\label{eq:rty}
{\cal V}\leq\max_{\Delta_0}\int\;dp\;P_{\rm in}(p)|\Omega(p)|
\left|\cos\frac{p\Delta_0}{\hbar}\right|=\int\;dp\;P_{\rm in}(p){\cal V}(p)\leq
1 .
\eeq
The generalized visibility yields the maximum ``distance" between
the intensities ${\cal N}_O$ and ${\cal N}_E$ and is bounded by
the ``local" visibility averaged over the momentum distribution of
the incoming state.

\begin{figure}[t]
\begin{center}
\includegraphics[width=8cm]{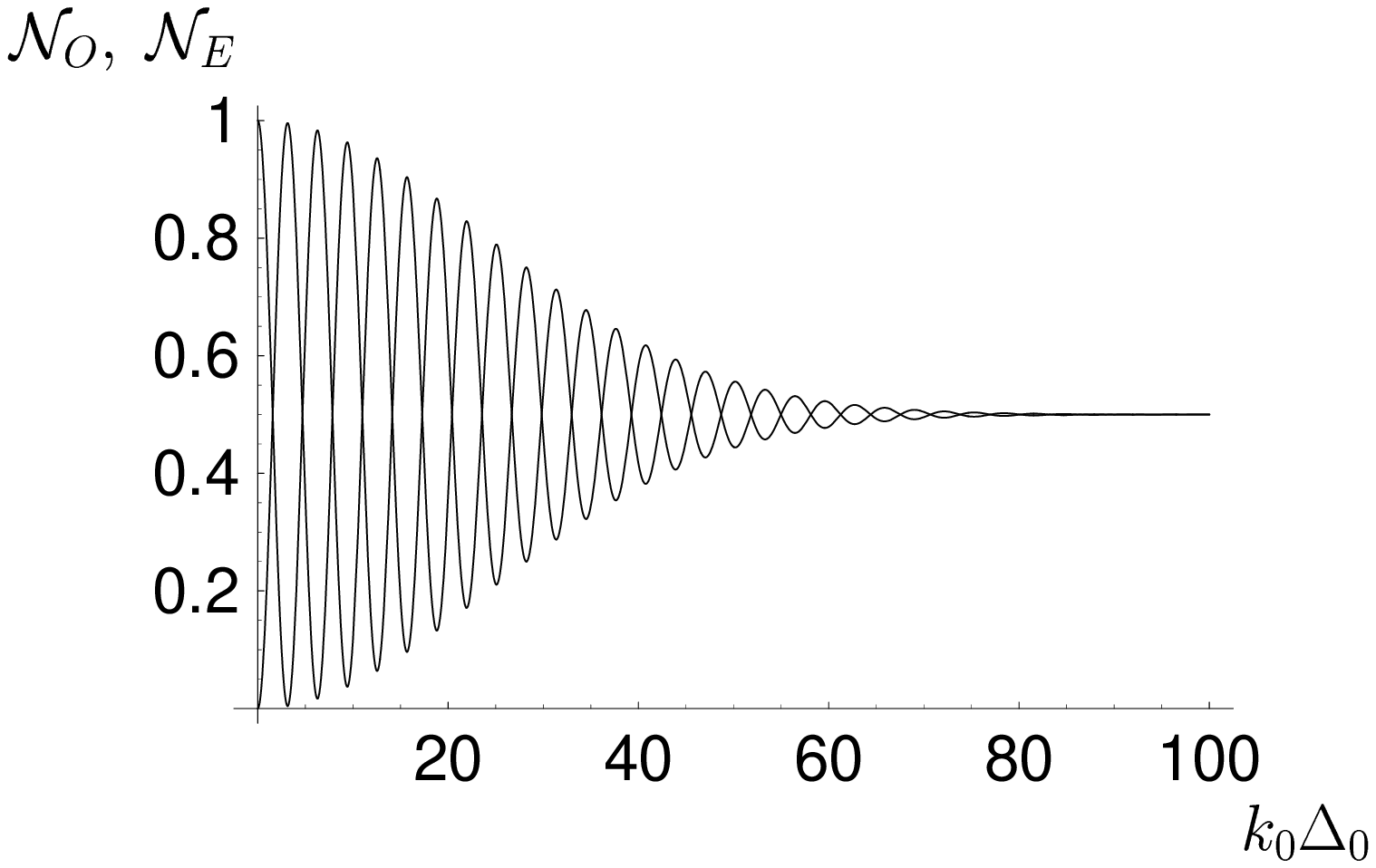}
\end{center}
\caption{${\cal N}_O$ and ${\cal N}_E$ versus $k_0\Delta_0$ ($k_0=p_0/\hbar$)
for an incoming Gaussian wave packet (\ref{eq:rum5}) with
$k_0\delta=12$ and no fluctuations. The two intensities differ in
phase by $\pi$ and their sum is 1. The generalized visibility
(\ref{eq:de0visb}) is 1.}
\label{fig:visi}
\end{figure}

For a fluctuation-free phase shifter, i.e.\ for
$w(\Delta)=\delta(\Delta)$, one obtains $\Omega(p)=1$ and the
generalized visibility (\ref{eq:de0visb}) becomes
\andy{nofluct}
\beq
\label{eq:nofluct}
{\cal V}=\max_{\Delta_0 }\left|\left\langle
\cos\frac{\hat p\Delta_0
}{\hbar }\right\rangle\right| =\max_{\Delta_0
}\left|\int\;dp\;P_{\rm in}(p)
\cos\frac{p\Delta_0}{\hbar }\right| =
\int\;dp\;P_{\rm in}(p)=1,
\eeq
for any incoming state $P_{\rm in}$. The example of an incoming
Gaussian wave packet
\andy{rum5}
\beq\label{eq:rum5}
P_{\rm in}(p)=\sqrt{\frac{2\delta^2}{\hbar^2\pi}
}\exp\left(-\frac{2\delta^2}{\hbar^2}(p-p_0)^2\right)
\eeq
is shown in Figure \ref{fig:visi}, where it is apparent that
${\cal V}=1$.

If, on the other hand, the phase shifter fluctuates, the amplitude
of the envelope function decreases and ${\cal V}<1$. We therefore
give an operational definition of decoherence, by means of a {\em
decoherence parameter}:
\andy{decimp}
\beq\label{eq:decimp}
\varepsilon\equiv1-{\cal
V}=1-\max_{\Delta_0}\left|\left\langle\Omega(\hat p)\cos\frac{\hat
p\Delta_0 }{\hbar }\right\rangle\right|.
\eeq
Notice that, by Eq.\ (\ref{eq:nofluct}), $\varepsilon=0$ for a
fluctuation-free phase shifter (quantum coherence perfectly
preserved), while $\varepsilon \to 1$ when the magnitude of the
fluctuations increases, $\Omega(p)\to0$ and the envelope function
in Figure \ref{fig:visi} squeezes out all oscillations, eventually
yielding ${\cal N}_O (\Delta_0)={\cal N}_E(\Delta_0)$,
independently of $\Delta_0$. Observe also that ${\cal V}$ and
$\varepsilon$ are independent of the coherence of the initial
state (namely, they do not depend on the off-diagonal terms of the
density matrix). In this sense they measure the {\em loss} of
quantum coherence.

\section{Examples}
\begin{figure}[t]
\begin{center}
\includegraphics[width=6.5cm]{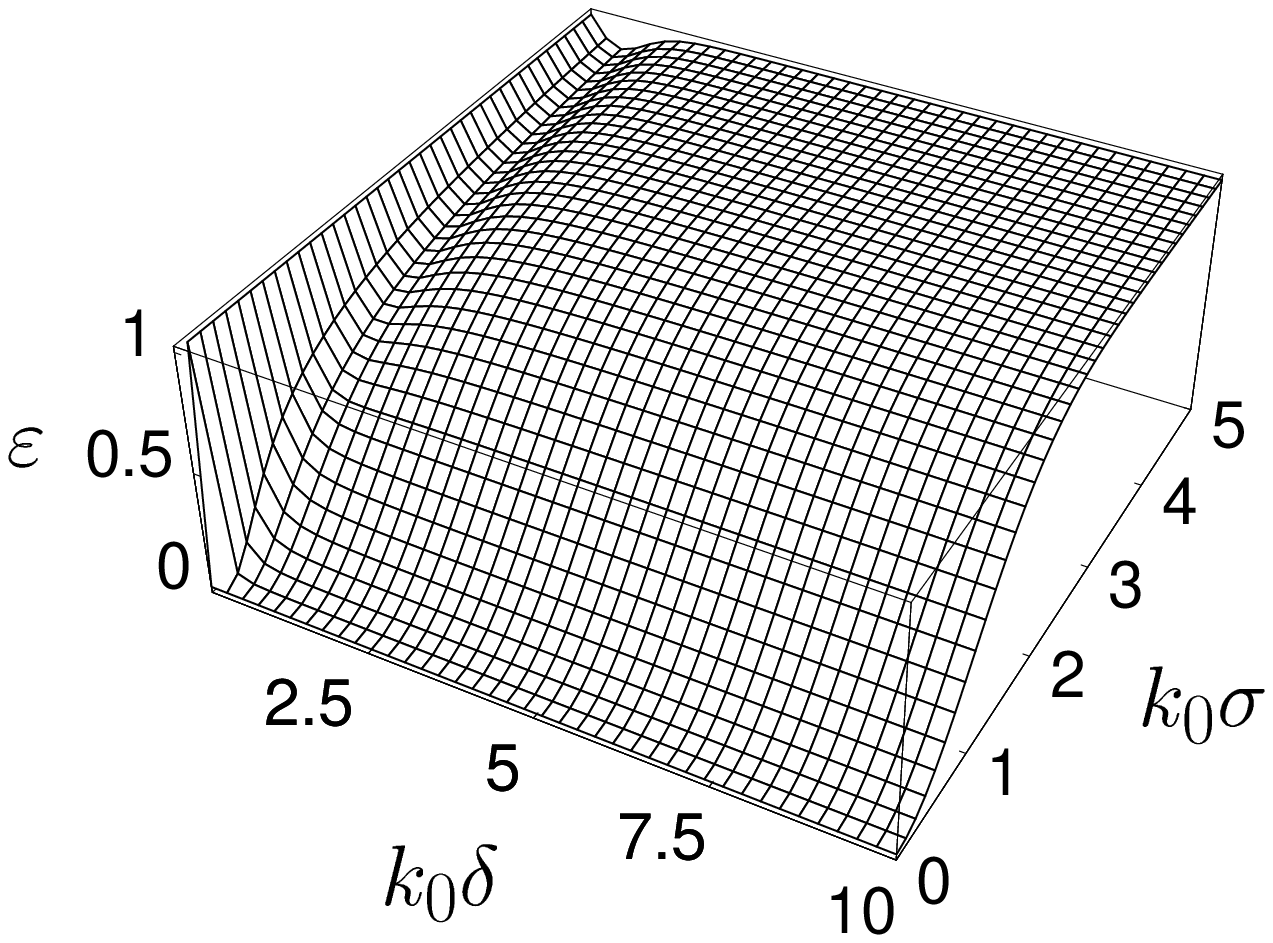}
\end{center}
\caption{Decoherence parameter $\varepsilon$
(\ref{eq:rum6}) versus width $\delta$ of the Gaussian wave packet
and standard deviation $\sigma$ of the fluctuating shifts
($k_0=p_0/\hbar$).}
\label{fig:gaun}
\end{figure}

\andy{sec-capa} \label{sec-capa}
Let us now look at some particular cases of fluctuations. Let the
phases be distributed according to a Gaussian law with standard
deviation $\sigma$
\andy{rum1}
\beq\label{eq:rum1}
w(\Delta)=\frac{1}{\sqrt{2\pi\sigma^2} }\exp\left(-\frac{\Delta^2
}{2\sigma^2 }\right),
\eeq
so that $\Omega(p)=\exp\left(-\frac{p^2\sigma^2}{2\hbar^2}\right)$
and the decoherence parameter reads
\andy{rum3}
\beq\label{eq:rum3}
\varepsilon=1-\max_{\Delta_0 }\left|\int\;dp\;P_{\rm in}(p)
\exp\left(-\frac{p^2\sigma^2}{2\hbar^2}\right)\cos
\left(\frac{p\Delta_0}{\hbar}\right)\right| .
\eeq
For Gaussian wave packets (\ref{eq:rum5}) one gets
\andy{rum6}
\beq\label{eq:rum6}
\varepsilon=1-\sqrt{\frac{\delta^2}{\delta^2+\sigma^2/4}
}\exp\left(-\frac{\delta^2}
{\delta^2+\sigma^2/4}\frac{\sigma^2k_0^2}{2}\right),
\eeq
with $k_0=p_0/\hbar$. In this case, as it is clear from Figure
\ref{fig:gaun}, at fixed $\delta$
the decoherence parameter (\ref{eq:decimp}) increases with
$\sigma$, although the details of its behavior are strongly
dependent on the spatial width of the packet $\delta$. This
behavior is in agreement with expectation: decoherence
$\varepsilon$ increases with the magnitude of fluctuations
$\sigma$.

\begin{figure}[t]
\begin{center}
\includegraphics[width=\textwidth]{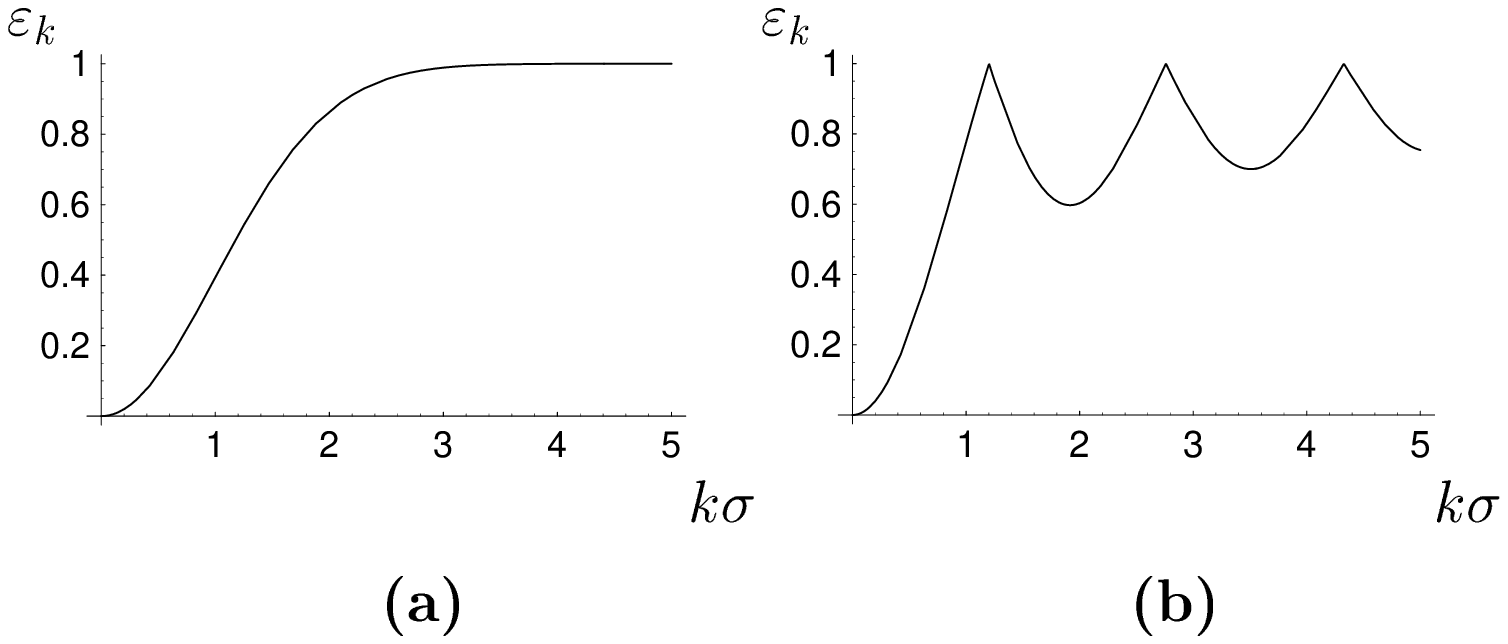}
\end{center}
\caption{(a) Decoherence parameter
 $\varepsilon_k$ (\ref{eq:rum4}) versus $k\sigma$,
 for a plane wave interacting with a
 shifter fluctuating according to (\ref{eq:rum1});
 (b) Decoherence parameter
 $\varepsilon_k$ (\ref{eq:dec2}) versus $k\sigma$,
 for a plane wave interacting with
 a shifter fluctuating according to (\ref{eq:sindis}).}
\label{fig:paga}
\end{figure}

For plane waves [$P_{\rm in}(p')=\delta(p-p')$]
\andy{rum4}
\beq\label{eq:rum4}
\varepsilon_k=1-\max_{\Delta_0 }\left|\int\;dp'\;\delta(p-p')
e^{-\frac{p'^2\sigma^2}{2\hbar^2}}\cos\left(\frac{p'\Delta_0}{\hbar}\right)\right|=1-
e^{-\frac{k^2\sigma^2}{2}},
\eeq
with $k=p/\hbar$. This is shown in Figure \ref{fig:paga}(a) and
can be obtained from (\ref{eq:rum6}) in the $\delta \to \infty$
limit. Notice that high momenta are more fragile against
fluctuations \cite{RS}.

Let now the phase shifts be distributed according to the law
\cite{fibonacci}
\andy{sindis}
\beq\label{eq:sindis}
w(\Delta)=\frac{1}{\pi}\frac{1}{\sqrt{4\sigma^2-\Delta^2} }.
\eeq
This is convenient from an experimental perspective and follows
from a phase $\Delta(t)=2\sigma\sin t$, where $t$ (``time") is a
parameter, uniformly distributed between 0 and $2\pi$. From
(\ref{eq:imptr0}) and (\ref{eq:wtilde})
\andy{lewa}
\barr\label{eq:lewa}
\Omega(p)=\int_{-2\sigma }^{2\sigma}\;
\frac{d\Delta}{\pi}\frac{e^{i\frac{p\Delta}{\hbar}}}
{\sqrt{4\sigma^2-\Delta^2} } =\int_{-\pi/2}^{\pi/2
}\frac{dt}{\pi}\exp\left(i\frac{2p\sigma}{\hbar}\sin t\right)=
J_0\left(\frac{2p\sigma}{\hbar}\right),
\earr
where $J_0$ is the Bessel function of order zero. The decoherence
parameter (\ref{eq:decimp}) reads
\andy{dec1}
\beq\label{eq:dec1}
\varepsilon=1-\max_{\Delta_0}\left|\int\;dp\;P_{\rm in}(p)
J_0\left(\frac{2p\sigma}{\hbar}\right)\cos\left(\frac{p\Delta_0}{\hbar}\right)\right|
\eeq
and for plane waves one obtains ($k=p/\hbar$)
\andy{dec2}
\beq\label{eq:dec2}
\varepsilon_{k}=1-\max_{\Delta_0}\left|J_0\left(\frac{2p\sigma}{\hbar}\right)
\cos\left(\frac{p\Delta_0}{\hbar}\right)\right|=1-|J_0(2k\sigma)|.
\eeq
This function is shown in \ref{fig:paga}(b): observe that
decoherence is {\em not} a monotonic function of the noise
$\sigma$ in (\ref{eq:sindis}).

A comparison between Figures \ref{fig:paga}(a) and
\ref{fig:paga}(b) is interesting. In both cases one observes
fragility at high momenta $p=\hbar k$. On the other hand, the
behavior of decoherence in Figure
\ref{fig:paga}(b) is somewhat anomalous and against naive
expectation. For a given $k$, there are situations where
decoherence $\varepsilon$ {\em decreases by increasing} the size
of fluctuations $\sigma$. Note also that we are considering
incoming plane waves, whence, according to (\ref{eq:qwe}),
$\epsilon_k=1-{\cal V}(p)$ and the decoherence parameter is
strictly related to the standard visibility of the interference
pattern. Therefore, in the anomalous regions, one observes an
increase in visibility by increasing the fluctuations of the phase
shifter, a phenomenon similar to stochastic resonance
\cite{stochastic}. However, this is true not only for plane waves, but also
for narrow packets in momentum space.

These results are related to well known phenomena in the classical
theory of partially coherent light
\cite{BornWolf}, where the visibility is expressed as the Fourier transform of
the spectral distribution of an incoherent source.

\section{Conclusions}

We have introduced and discussed a decoherence parameter defined
in terms of a generalized visibility of the interference pattern
in a double-slit experiment (MZI). Although the notion of
visibility is not directly related to that of decoherence (see
post-selection experiments \cite{postsel}) our results corroborate
the ideas expressed in \cite{fibonacci} and make it obvious that
the concept of ``loss" of quantum mechanical coherence deserves
clarification and additional investigation.

It would also be interesting to discuss analogies and differences
with conceptual experiments in which decoherence is complemented
by Welcher-Weg information \cite{englert}.

\end{document}